\documentclass[11pt]{imsart}

\RequirePackage{amsthm,amsmath,amsfonts,amssymb}
\RequirePackage[authoryear]{natbib}
\RequirePackage[colorlinks,citecolor=blue,urlcolor=blue]{hyperref}
\RequirePackage{graphicx}

\usepackage[margin=3cm]{geometry}
\usepackage{lipsum}

\usepackage{bm}

\startlocaldefs

\theoremstyle{remark}

\endlocaldefs

\newcommand{\nmathbf}{\bm}

\newcommand*{\mc}[1]{\multicolumn{1}{|c|}{#1}}

\def\bfV{\nmathbf V}

\def\bfX{\nmathbf X}

\def\bfbeta   {\nmathbf \beta}

\def\bftheta  {\nmathbf \theta}

\def\bfSigma  {\nmathbf \Sigma}

\def\bfPhi    {\nmathbf \Phi}

\def\boldfacefake#1{\kern-4pt
   \hbox{ \mathsurround=0pt
   \hbox to 0.4pt{$#1$\hss}\hbox to 0.4pt{$#1$\hss}\hbox {$#1$}}}

\newcommand{\btable}{\begin{table}[h]\centering}
\newcommand{\etable}{\end{table}}
\newcommand{\bt}{\begin{parag}\small \let\b=\nsb \let\sb=\nssb \begin{tabular}}
\newcommand{\et}{\end{tabular}\let\b=\nb \let\sb=\nsb\end{parag}}

\newenvironment{parag}{\par}{\par}

\newcommand{\be}{\begin{eqnarray}}
\newcommand{\ee}{\end{eqnarray}}
\newcommand{\ba}{\begin{eqnarray*}}
\newcommand{\ea}{\end{eqnarray*}}

\newcommand{\reals}{\mbox{\rm I\kern-.20em R}}
\newcommand{\sreals}{\mbox{\small \rm I\kern-.20em R}}

\begin{document}

\thispagestyle{empty}
\pagestyle{empty}

\begin{frontmatter}
\title{\LARGE Measuring performance for end-of-life care}
\runtitle{Measuring performance for end-of-life care}

\begin{aug}
\small
\author[A]{\fnms{Sebastien} \snm{Haneuse}\ead[label=e1]{}},
\author[B]{\fnms{Deborah} \snm{Schrag}\ead[label=e2]{}},
\author[A]{\fnms{Francesca} \snm{Dominici}\ead[label=e3]{}},
\author[C]{\fnms{Sharon-Lise} \snm{Normand}\ead[label=e4]{}},
\and
\author[A,D,E]{\fnms{Kyu Ha} \snm{Lee}\ead[label=e5]{}}
\address[A]{Department of Biostatistics, Harvard T.H. Chan School of Public Health}

\address[B]{Division of Population Sciences, Dana-Farber Cancer Institute}

\address[C]{Department of Health Care Policy, Harvard Medical School}

\address[D]{Department of Nutrition, Harvard T.H. Chan School of Public Health}

\address[E]{Department of Epidemiology, Harvard T.H. Chan School of Public Health}

\end{aug}

\begin{abstract}
Although not without controversy, readmission is entrenched as a hospital quality metric, with statistical analyses generally based on fitting a logistic-Normal generalized linear mixed model. Such analyses, however, ignore death as a competing risk, although doing so for clinical conditions with high mortality can have profound effects; a hospitals seemingly good performance for readmission may be an artifact of it having poor performance for mortality. In this paper we propose novel multivariate hospital-level performance measures for readmission and mortality, that derive from framing the analysis as one of cluster-correlated semi-competing risks data. We also consider a number of profiling-related goals, including the identification of extreme performers and a bivariate classification of whether the hospital has higher-/lower-than-expected readmission and mortality rates, via a Bayesian decision-theoretic approach that characterizes hospitals on the basis of minimizing the posterior expected loss for an appropriate loss function. In some settings, particularly if the number of hospitals is large, the computational burden may be prohibitive. To resolve this, we propose a series of analysis strategies that will be useful in practice. Throughout the methods are illustrated with data from CMS on $N$=17,685 patients diagnosed with pancreatic cancer between 2000-2012 at one of $J$=264 hospitals in California.
\end{abstract}


\end{frontmatter}

\section{Introduction}
\label{sec:intro}

The profiling and ranking of institutions is a major societal endeavor. Aimed at improving the quality of the services that institutions provide, two major areas where profiling and ranking are a matter of public policy are education~\citep{goldstein1996league, leckie2009limitations, bates2019education} and health care. For the latter, readmission rates have become a central tool in assessing variation in quality of care globally~\citep{westert2002international, kristensen2015roadmap}, including specific efforts in England~\citep{friebel2018national}, Scotland~\citep{HNSNSS2019hospital}, Denmark~\citep{ridgeway2019benchmarking}, and Canada~\citep{samsky2019trends}. In the United States, the Hospital Inpatient Quality Improvement Program, established by the Medicare Prescription Drug, Improvement and Modernization Act of 2003, for example, provides financial incentives for hospitals to publicly report 30-day all-cause readmission and all-cause mortality rates for acute myocardial infarction, heart failure, and pneumonia. More recently, the Affordable Care Act of 2012 established the Hospital Readmissions Reduction Program which instructs the Centers for Medicare and Medicaid Services (CMS; a federal agency charged with administering and facilitating the administration of health care to the poor and the elderly) to tie hospital reimbursement rates to excess readmissions for these three conditions, as well as for chronic obstructive pulmonary disorder, elective total hip and/or knee replacement surgery and coronary artery bypass graft surgery. Whether a particular hospital has `excess' readmissions is quantified through the so-called \textit{excess readmission ratio}, the calculation of which is currently based on the fit of a logistic-Normal generalized linear mixed model (GLMM) to the binary outcome of whether a readmission occurred within some time interval~\citep{normand1997statistical, normand2016league}.

Common to each of the conditions currently evaluated by CMS is that post-diagnosis prognosis for patients is good and mortality low. As such, whether there is variation across hospitals in mortality may not have a large impact on conclusions regarding readmission. This may not be the case, however, for conditions for which prognosis is poor, mortality high, and the clinical management of patients is focused on end-of-life palliative care. One such condition is pancreatic cancer for which an estimated 56,770 new cases were diagnosed in the U.S. in 2019, and for which 5-year survival is estimated to only be 9.3\% (https://seer.cancer.gov/statfacts/html/pancreas.html). If interest lies in understanding variation of performance of end-of-life care for pancreatic cancer and other terminal conditions, as well as in developing policies to improve quality, then application of the current approach that ignores death has the potential to have profound effects. In particular, that a hospital seemingly has good performance for readmission may be an artifact of it having poor performance for mortality.

Based on these considerations a novel conceptualization of excess readmissions that explicitly accounts for mortality is needed. Towards this, we propose a new general framework for measuring hospital performance for end-of-life care. The framework has four key components, the first of which is to embed the joint analysis of readmission and mortality with a recently-proposed Bayesian modeling framework for cluster-correlated semi-competing risks data~\citep{lee2016hierarchical}. The remaining three components of the proposed framework constitute the contribution of this paper. Specifically, we first propose two novel metrics for assessing quality of end-of-life care: the \textit{cumulative excess readmission ratio} and the \textit{cumulative excess mortality ratio}. As we elaborate upon, these metrics are `cumulative' in the sense that they consider events up to some pre-specified time point (e.g. 30 or 90 days). The second is motivated by the notion that policies for monitoring and improving quality of end-of-life care should be developed on the basis of simultaneous consideration of a hospitals excess readmissions and excess deaths~\citep{haneuse2018assessment}. For example, quality improvement policies may be tailored to whether a hospital has higher-than expected or lower-than expected readmissions simultaneously with whether it has higher-than expected or lower-than expected mortality (i.e. according to which of four categories that it falls into). To facilitate this, the second component of the framework is a novel decision-theoretic loss function-based approach to hospital profiling jointly on the basis of readmission and mortality. While there is a modest but rich statistical literature on loss-function based profiling, it has generally focused on settings where the outcome of interest is univariate~\citep{Lair:Loui:1989, Shen:Loui:1998, Lin:Loui:2006, paddock2006flexible, ohlssen2007flexible, Lin:Loui:2009, Padd:Loui:2011, paddock2014statistical, hatfield2017regulator}. Moreover, while there has been work on novel multivariate hierarchical models that consider a range of performance measures simultaneously~\citep{landrum2003selection, daniels2005longitudinal, robinson2006hierarchical}, to the best of our knowledge, no-one has considered loss-function based joint profiling, in particular for the end-of-life contexts we are interested in. The third contribution of this paper is a series of computational strategies together with software in the form of the \texttt{SemiCompRisks} package for \texttt{R}. Central to this component are the use of Gauss-Hermite and Gauss-Legendre integration to approximate multivariate integrals required in the calculation of the proposed cumulative excess readmission and mortality ratios, as well as a series of practical approaches to mitigating computational burden associated with finding the minimizer of the Bayes risk in the proposed loss function-based approach to profiling. Throughout this paper, key concepts and proposed methods are illustrated using data from an on-going study of between-hospital variation in quality of end-of-life care for patients diagnosed with pancreatic cancer (described next). Where appropriate, detailed derivations and additional results are provided in the Supplementary Materials.

\section{Post-discharge outcomes among patients diagnosed with pancreatic cancer}
\label{sec:data}

We consider data from CMS on $N$=17,685 patients aged 65 years or older who received a diagnosis of pancreatic cancer during a hospitalization that occurred between 2000-2012 at one of $J$=264 hospitals in California with at least 10 such patients, and were discharged alive. Table SM-1 in the Supplementary Materials presents a summary of select patient-specific covariates, as well as joint outcomes of readmission and mortality at 90 days post-discharge within levels of those covariates; see \citet{lee2016hierarchical} for details on the rationale for a 90-day window. Note, for the purposes of this paper we only consider the first readmission following discharge from the index hospitalization.

Overall, 16.0\% of the patients in the sample (2,835) were readmitted and subsequently died within 90 days of discharge; 18.5\% (3,268) were readmitted within 90 days but did not die (i.e. were censored for death at 90 days); 35.2\% (6,321) died within 90 days without experiencing a readmission event; and, 30.3\% (5,351) were censored at 90 days without having experienced either a readmission event or death. Thus, more than 50\% of patients who were discharged alive went on to die within 90 days.

\begin{figure}[h!]
	\centering
	\includegraphics[width=3.5in]{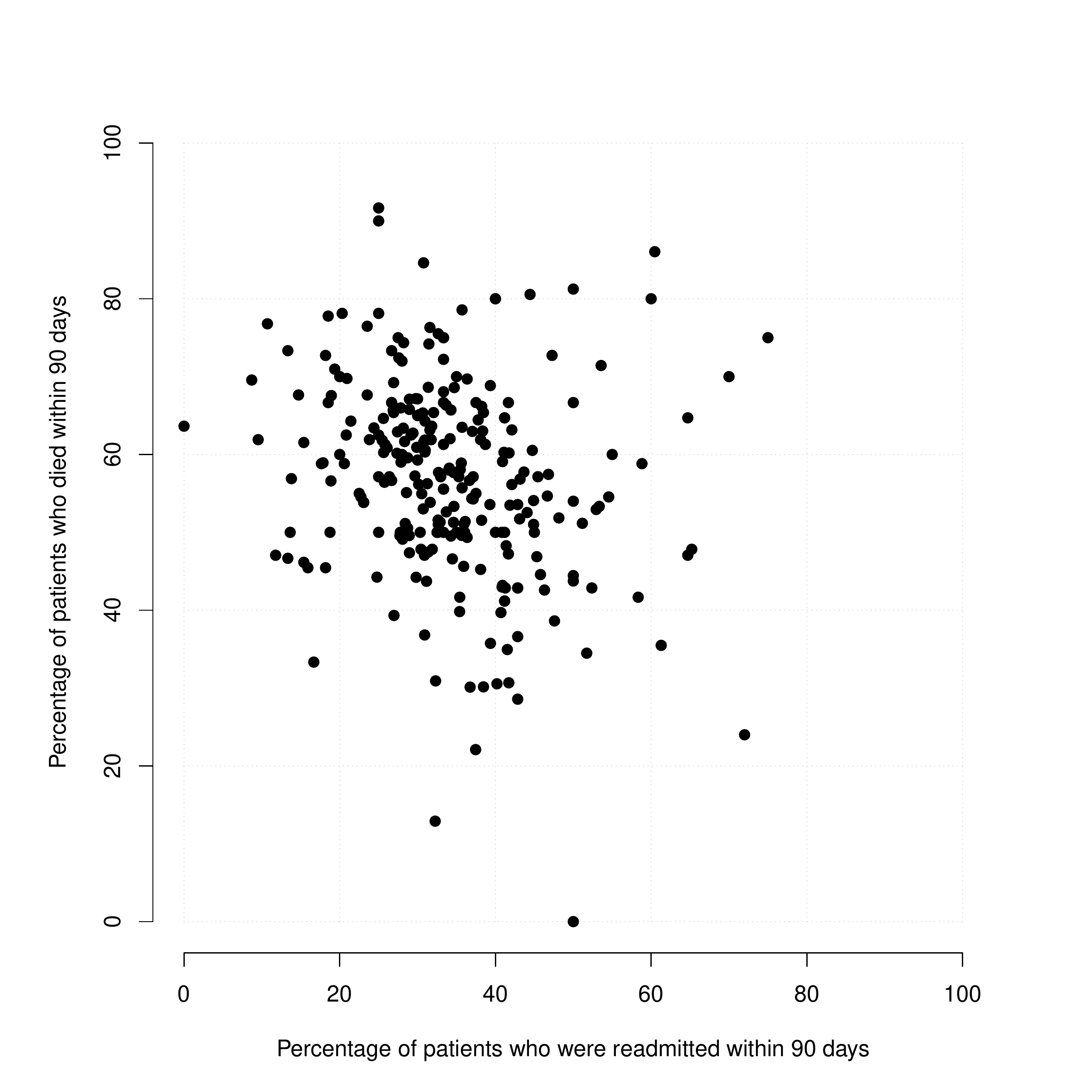}
	\caption{\label{fig:PC:observedY} Marginal 90-day readmission and 90-day mortality rates across $J$=264 hospitals in California with at least 10 patients aged 65 years or older and diagnosed with pancreatic cancer between 2000-2012. Note, the rates are marginal in the sense that they are not covariate-adjusted.} 
\end{figure}

Figure \ref{fig:PC:observedY} presents a scatterplot of the percentage of patients who were readmitted within 90 days (marginalized over death) and the percentage of patients who died within 90 days (marginalized over readmission) across the $J$=264 hospitals. Noting that neither quantities are covariate-adjusted, we find that there is substantial variation in both the marginal 90-day readmission rate and the marginal 90-day mortality rate across the $J$=264 hospitals, and that there is an indication that performance with respect to readmission is negatively correlated to that with respect to mortality (estimated Pearson correlation of -0.21).

Finally, we note that there is substantial variation in the patient case-mix across the hospitals. The inter-quartile range for the percentage non-white, for example, is 10.5-35.4\%, while the inter-quartile range for the percentage of patients who are discharged to a hospice, skilled nursing facility or intensive care facility is 24.1-38.9\%; see Figures SM-1 and SM-2 in the Supplementary Materials.

\section{Profiling for binary outcomes}
\label{sec:binary}

Let $Y^*_{ji}$ = 0/1 be a binary indicator of whether or not the $i^{th}$ patient in the $j^{th}$ hospital was readmitted within 90 days of discharge. Note, if a patient died prior to readmission within 90 days their outcome would be set to $Y^*_{ji}$ = 0. An analysis could proceed on the basis of a logistic-Normal GLMM:
\be
\label{model:ln:glmm}
	\mbox{logit} \Pr(Y^*_{ji} = 1| \bfX^*_{ji}; \bfbeta^*, V^*_j)\ =\ \bfX^{*T}_{ji}\bfbeta^*\ +\ V^*_j
\ee
\noindent where $\bfX^*_{ji}$ is a vector of patient-specific covariates measured prior to discharge and $V^*_j$ is a hospital-specific random effect that is taken to arise from a Normal(0, $\sigma_v^2$) distribution.

\subsection{A measures for performance}
\label{sec:binary:measures}

Given model (\ref{model:ln:glmm}), \cite{normand1997statistical} define the hospital-specific \textit{adjusted outcome rate}:
\[
	\mu_j^a\ =\ \frac{1}{n_j}\sum_{i=1}^{n_j}\Pr(Y^*_{ji} = 1| \bfX^*_{ji}; \bfbeta^*, V^*_j)
\]
and the \textit{standardized adjusted outcome rate}:
\[
	\mu_j^s\ =\ \frac{1}{n_j}\sum_{i=1}^{n_j} \mbox{E}[\Pr(Y^*_{ji} = 1| \bfX^*_{ji}; \bfbeta^*, V^*)],
\]
where the expectation in $\mu_j^s$ is with respect to the Normal(0, $\sigma_v^2$) distribution for $V^*$. Based on these, the \textit{excess readmission ratio} is $\theta_j$ = $\mu_j^a/\mu_j^s$. Intuitively, $\theta_j$ represents the extent to which the `observed' readmission rate for the $j^{th}$ hospital differs from the `expected' rate for the specific number and case-mix of patients actually treated at the hospital~\citep{normand2016league}. If $\theta_j$ $>$ 1.0, one concludes that the rate was higher than would be expected, given the patient case-mix, indicating poor performance. If $\theta_j$ $<$ 1.0, one can conclude that the rate was lower than expected, given the patient case-mix, indicating good performance.

Practically, if estimation and inference is to be performed within the frequentist paradigm then $\mu_j^a$ can be estimated by plugging in point estimates for $V^*_j$ and $\bfbeta^*$. Furthermore, $\mu_j^s$ can be estimated by approximating the expectation using Gauss-Hermite quadrature based on estimates of $\bfbeta^*$ and the variance component $\sigma_v^2$~\citep{stoer2013introduction}. If estimation and inference is to be performed via MCMC within the Bayesian paradigm, then posterior samples of $\theta_j$ can be obtained by calculating $\mu_j^a$ and $\mu_j^s$ at their current values in the MCMC scheme, again using Gauss-Hermite quadrature for the latter.


\subsection{Application to pancreatic cancer data}
\label{sec:binary:application}

Returning to the CMS pancreatic cancer data, we performed a Bayesian analyses for the binary indicator of 90-day readmission and (separately) the binary indicator of 90-day mortality, each based on model (\ref{model:ln:glmm}) with the components of $\bfX^*$: sex, age, race, admission route, Charlson-Deyo comorbidity score~\citep{deyo1992adapting}, length of stay and discharge location.

\begin{table}[h!]
\caption{\label{tab:table-2} Posterior medians for the odds ratio (OR) parameters from separate logistic-Normal generalized linear mixed models (LN-GLMM) for 90-day readmission and 90-day mortality, and for the hazard ratio (HR) parameters from a PEM-MVN hierarchical semi-competing risks model; see Sections \ref{sec:binary} and \ref{sec:clusteredSCR:model} for additional detail. Further detail, together with 95\% credible intervals (CI), is provided in the Supplementary Materials. Note, estimates highlighted in boldface have 95\% CIs that exclude 1.0.}
\centering
\scalebox{1}{%
\begin{tabular}{lccccccc}
\hline
					&& \multicolumn{2}{c}{\textbf{LN-GLMM}} && \multicolumn{3}{c}{\textbf{PEM-MVN}}\\
\cline{3-4}\cline{6-8}
  					&& Readmission 	& Death 		&& Readmission & Death prior 		& Death after \\
					&&				&			&& prior to death & to readmission	& readmission \\
					&& OR 			& OR 		&& HR & HR & HR \\ 
\hline
\textbf{Sex}: Female				&& \bf{0.90}	& \bf{0.81}		&& \bf{0.85}	& \bf{0.80} & \bf{0.85} \\ 
\textbf{Age$^\dag$}				&& \bf{0.86}	& \bf{1.16}		&& \bf{0.90}	& \bf{1.10} & \bf{1.06} \\ 
\textbf{Race}: Non-white			&& \bf{1.43}	& 0.94		&& \bf{1.26}	& \bf{0.84} & 1.06 \\ 
\textbf{Admission}: Other			&& 1.06 		& \bf{1.34}		&& \bf{1.15}	& \bf{1.29} & \bf{1.12} \\ 
\textbf{Charlson-Deyo score}: $>1$ 	&& \bf{1.18}	& \bf{1.39} 	&& \bf{1.26}	& \bf{1.23} & \bf{1.23} \\ 
\textbf{Length of stay$^*$} 		&& 1.01		& \bf{0.86} 	&& 0.98		& \bf{0.90} & \bf{0.89} \\ 
\textbf{Discharge location$^+$} \\
~~Home with care 				&& \bf{0.88}	& \bf{1.87}		&& 1.06		& \bf{1.90} & \bf{1.53} \\ 
~~Hospice 					&& \bf{0.07}	& \bf{17.6} 	&& \bf{0.24}	& \bf{10.2} & \bf{3.21} \\ 
~~ICF/SNF	 				&& \bf{0.63}	& \bf{3.48} 	&& 1.00		& \bf{3.50} & \bf{2.07} \\ 
~~Other 						&& \bf{0.62}	& \bf{2.35} 	&& 0.88		& \bf{2.77} & \bf{1.39} \\
\hline
   	\multicolumn{8}{l}{\scriptsize $^\dag$ Standardized so that 0 corresponds to an age of 77 years and so that a one unit increment corresponds to 10 years}\\
   	\multicolumn{8}{l}{\scriptsize $^*$ Standardized so that 0 corresponds to 10 days and so that a one unit increment corresponds to 7 day} \\
   	\multicolumn{8}{l}{\scriptsize $^+$ Referent category is `Home without care'} \\
\end{tabular}}
\end{table}

\begin{figure}[!htbp]
	\centering
	\includegraphics[width=6in]{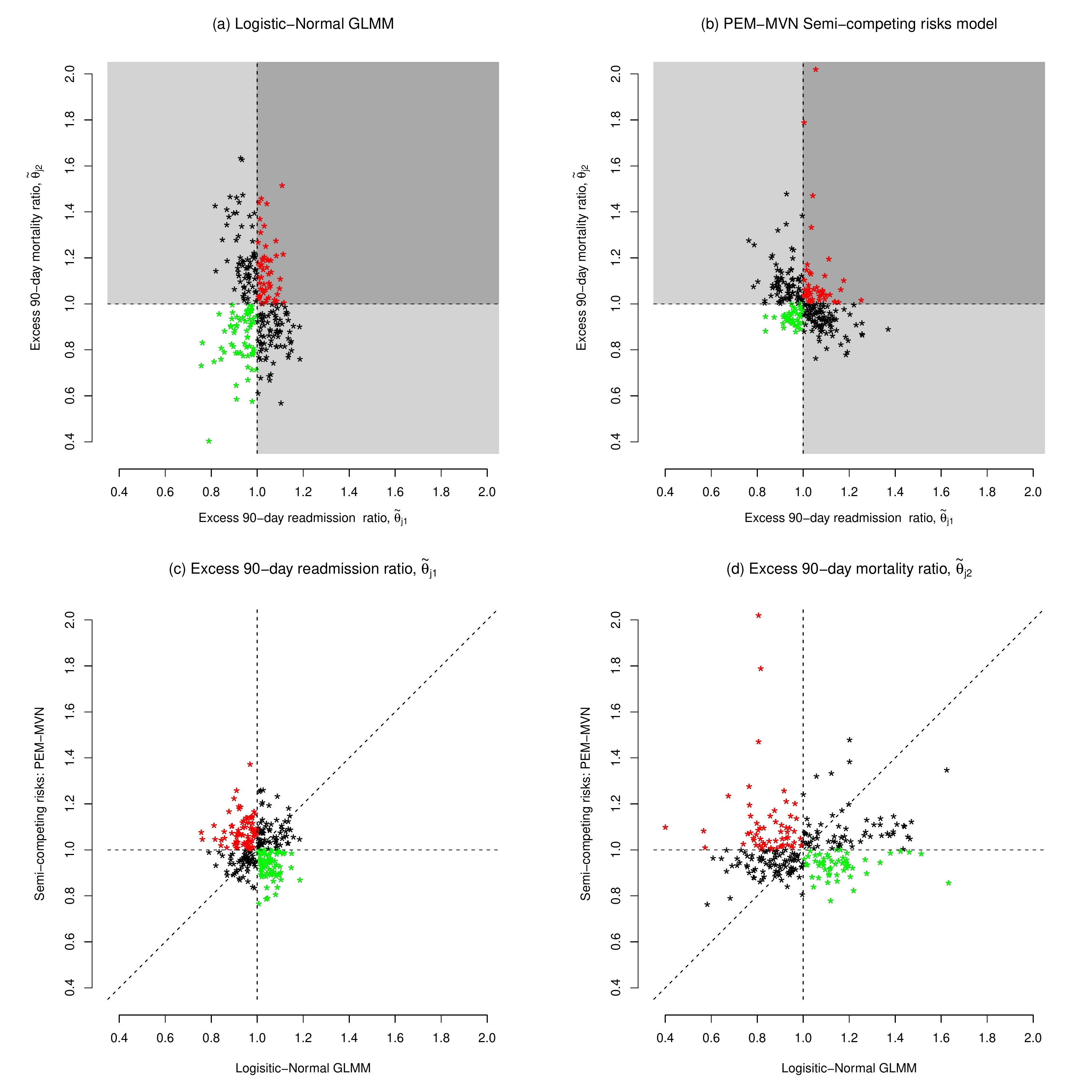}
	\caption{\label{fig:PC:joint} Excess 90-day readmission and 90-day mortality ratios across $J$=264 hospitals in California with at least 10 patients aged 65 years or older and diagnosed with pancreatic cancer between 2000-2012. Shown in panels (a) and (b) are posterior medians, color-coded by whether the value are less than or greater than 1.0: in (a) the results are based on a Bayesian fit of the logistic-Normal GLMM (see Section \ref{sec:binary}); in (b) the results are based on a PEM-MVN semi-competing risk model (see Sections \ref{sec:clusteredSCR} and \ref{sec:measures}). In panels (c) and (d), hospitals indicated with green dots were re-classified as having lower-than-expected readmission or mortality by the semi-competing risks analysis (i.e. benefitted), while those indicated with a red dot were re-classified as having higher-than-expected readmission or mortality (i.e. lost).} 
\end{figure}

The first two columns of Table \ref{tab:table-2} present posterior medians of the odds ratio parameters from the two logistic-Normal models (i.e. exp$\{\beta^*\}$), with boldface highlighting indicating that the corresponding central 95\% credible interval excluded 1.0; see Section SM-2 for additional detail. Additionally, Figure \ref{fig:PC:joint}(a) presents the posterior medians of the excess 90-day readmission ratio, denoted by $\widetilde{\theta}_{j1}$, and of the excess 90-day mortality ratio, denoted by $\widetilde{\theta}_{j2}$. Several aspects of the results are worth noting. First, there is substantially greater variation across the hospitals in performance for 90-day mortality (with $\widetilde{\theta}_{j2}$ varying between 0.40 to 1.63) than in performance for 90-day readmission ($\widetilde{\theta}_{j1}$ 0.76 to 1.19). Second, there is significant discordance in the classification of hospitals are being `good' or `poor' performers for the two outcomes (i.e. whether $\widetilde{\theta}_{j1}$ or $\widetilde{\theta}_{j2}$ is less than or greater than 1.0). Specifically, we find that 51 hospitals represented by red dots are classified as being poor performers with respect to both readmission and mortality, while 64 hospitals represented by green dots are classified as being good performers for both. The black dots indicate hospitals with mixed performance, with the 69 in the top-left quadrant classified as being good performers with respect readmission but poor performers with respect to mortality and 80 in the bottom-right quadrant classified as being poor performers with respect to readmission but good performers with respect to readmission.

Together Figures \ref{fig:PC:observedY} and \ref{fig:PC:joint} provide a reasonable basis for questioning the role that performance with respect to 90-day mortality plays when considering performance with respect to 90-day readmission. Specifically, as alluded to in the Introduction, it is plausible that some hospitals are being erroneously classified because mortality is not being explicitly accounted for in either the underlying regression analyses or in the performance metric.

\section{Bayesian analyses of cluster-correlated semi-competing risks data}
\label{sec:clusteredSCR}

As will become clear, the metrics and methods proposed in Sections \ref{sec:measures} and \ref{sec:goals} follow from the fit of a hierarchical model for the cluster-correlated semi-competing risks data. In this section we review a framework for such models proposed by \citet{lee2016hierarchical}. Throughout, while the methods are applicable to any cluster-correlated semi-competing risks data setting, we use the terms `hospital' and `patient' to be inline with the data application.

\subsection{A hierarchical illness-death model}
\label{sec:clusteredSCR:model}

Let $J$ denote the number of hospitals and $n_j$ the number of patients in the $j^{th}$ hospital, $j=1, \ldots, J$. Let $T_{ji1}$ and $T_{ji2}$ denote the times to readmission and death, respectively, for the $i^{th}$ patient in the $j^{th}$ hospital, respectively, for $i = 1, \ldots, n_j$ and $j = 1, \ldots, J$. In an illness-death model, the rates at which a given patient transitions between the initial state (i.e. discharged alive) and the state of experiencing a readmission event and/or the state of experiencing a death event are assumed to be governed by three hazard functions: $h_1(t_1)$, the cause-specific hazard for readmission given that a mortality event has not occurred; $h_2(t_2)$, the cause-specific hazard for mortality given that a readmission event has not occurred; and $h_3(t_2|t_1)$, the hazard for mortality given than a readmission event occurred at time $t_1$. Towards the analysis of cluster-correlated semi-competing risks data, \cite{lee2016hierarchical} proposed the following hierarchical illness-death model:
\be
\label{HID:h1}
	h_1(t_{ji1}| \bfX_{ji1}; \gamma_{ji}, h_{01}, \bfbeta_1, V_{j1}) &=& \gamma_{ji}\ h_{01}(t_{ji1})\ \mbox{exp}\{\bfX_{ji1}^T\bfbeta_1\ +\ V_{j1}\}, \\
\label{HID:h2}
	h_2(t_{ji2}| \bfX_{ji2}; \gamma_{ji}, h_{02}, \bfbeta_2, V_{j2}) &=& \gamma_{ji}\ h_{02}(t_{ji2})\ \mbox{exp}\{\bfX_{ji2}^T\bfbeta_2\ +\ V_{j2}\}, \\
\label{HID:h3}
	h_3(t_{ji2}|t_{ji1}, \bfX_{ji3}; \gamma_{ji}, h_{03}, \bfbeta_3, V_{j3}) &=&\gamma_{ji} \ h_{03}(t_{ji2}| t_{ji1})\ \mbox{exp}\{\bfX_{ji3}^T\bfbeta_3\ +\ V_{j3}\},
\ee
where $\gamma_{ji}$ is a patient-specific frailty, $\bfX_{jig}$ is a vector of transition/patient-specific covariates, $\bfbeta_g$ is a vector of transition-specific fixed-effect log-hazard ratio regression parameters and $\bfV_j = (V_{j1}, V_{j2}, V_{j3})$ is a vector of cluster-specific random effects. For the patient-specific $\gamma_{ji}$ frailties, consistent with much of the literature on illness-death models, \cite{lee2016hierarchical} proposed that they be assumed to arise from a common Gamma($\theta^{-1}$, $\theta^{-1}$) distribution, such that $\mbox{E}[\gamma_{ji}] = 1$ and $\mbox{V}[\gamma_{ji}] = \theta$. For the transition-specific baseline hazard functions, we first note that because $h_{03}(t_{ji2}| t_{ji1})$ in expression (\ref{HID:h3}) is conditional on the timing of the non-terminal event, analysts are, in principle, faced with the task of specifying its functional form all possible $t_1$. Practically, this will likely be an onerous task and is typically mitigated via the adoption of a Markov model, such that $h_{03}(t_{ji2}| t_{ji1}) = h_{03}(t_{ji2})$, or a semi-Markov model, such that $h_{03}(t_{ji2}| t_{ji1}) =  h_{03}(t_{ji2} - t_{ji1})$~\citep[e.g.][]{Xu:Kalb:2010}. Whichever of these is adopted, \cite{lee2016hierarchical} proposed two specifications for the transition-specific baseline hazard functions: (i) that $h_{0g}(t)$ = $\alpha_{w,g} \kappa_{w,g} t^{\alpha_{w,g}-1}$, so that it corresponds to the hazard of a Weibull($\alpha_{w,g}$, $\kappa_{w,g}$) distribution; and (ii) that $h_{0g}$(t) be structured via a \textit{piecewise exponentials mixture} (PEM) model, with the mixing taking place over the number and placement of knots \citep{mckeague2000bayesian, sebastien2008separation}. Finally, \cite{lee2016hierarchical} proposed two specifications for the distribution of cluster-specific random effects, $\bfV_j$: (i) a mean-zero multivariate Normal distribution (MVN) with variance-covariance matrix $\bfSigma_V$; and, (ii) a Dirichlet process mixture (DPM) of MVNs \citep{ferguson1973bayesian, bush1996semiparametric, walker1997hierarchical}.

\subsection{Bayesian estimation and inference}
\label{sec:clusteredSCR:analyses}

To complete the Bayesian specification prior distributions and corresponding hyperparameters are needed. The exact nature of these depends on various choices made at the first stage of the model specification, including whether parametric Weibull or nonparametric PEM baseline hazard functions are adopted and whether a parametric or nonparametric specification of the distribution for the hospital-specific random effects, $\bfV_j$, is adopted; see \cite{lee2016hierarchical} for a detailed overview.

Given a complete specification of the model, samples from the joint posterior distribution can be obtained via a reversible-jump MCMC algorithm, implemented in the \texttt{SemiCompRisks} package for \texttt{R}, with model comparison based on the deviance information criterion~\citep[DIC;][]{spiegelhalter2002bayesian, millar2009comparison} and/or the log-pseudo marginal likelihood statistic~\citep[LPML;][]{geisser1993predictive}.

\subsection{Application to pancreatic cancer data}
\label{sec:clusteredSCR:application}

We fit four models to the CMS data, corresponding to combinations of the two baseline hazard specifications (i.e. Weibulls and PEM) and two specifications for the hospital-specific random effects (i.e. MVN and DPM). In each, we adopted semi-Markov specifications for $h_{03}(t_{ji2}| t_{ji1})$ and included the same set of covariates in $\bfX$ as those in $\bfX^*$ for the logistic-Normal models in Section \ref{sec:binary:application}, in each of the three transitions. Comprehensive details including the choice of hyperparameters, convergence of the MCMC schemes and posterior summaries are provided in Section SM-3 of the Supplementary Materials.

Based on the DIC and LMPL model comparison criteria, the model with a PEM specification for the baseline hazard functions and a MVN for the hospital-specific random effects (henceforth labelled as the `PEM-MVN model') was found to have the best fit (see Table SM-4 in the Supplementary Materials). The third, fourth and fifth columns of Table \ref{tab:table-2} present posterior medians for the hazard ratio parameters (i.e. exp$\{\beta_g\}$) for the three transitions, with boldface highlighting again indicating that the corresponding central 95\% credible interval excluded 1.0.

Comparing the first and third columns, we see that the two analyses indicate different predictor profiles for 90-day readmission. In particular, the semi-competing risks analysis indicates that patients who are admitted via some route other than the ER are at increased risk for 90-day readmisson whereas the standard LN-GLMM analysis indicated no such association. Furthermore, taking mortality into account dramatically changed the profile of associations between discharge location and risk of 90-day readmission.

Comparing the second column to the fourth and fifth we see that the set of predictors that have 95\% credible intervals that exclude 1.0 generally coincide between the two analyses. Furthermore, from the semi-competing risks analysis the associations between a given covariate and death are generally in the same direction whether death is being considered prior to or after a readmission event. The sole exception is for non-white race for which there is evidence of a decreased risk of death prior to a readmission event but no evidence of an association with death after a readmission event. 

\section{Performance metrics for end-of-life care}
\label{sec:measures}

Having reframed the investigation of variation in risk of readmission as a problem of semi-competing risks, we propose new metrics for performance for end-of-life care.

\subsection{Readmission}
\label{sec:measures:T1}

To characterize hospital performance with respect to readmission, we define the \textit{cumulative excess readmission ratio} for $t_1 > 0$ as:
\be
\label{eqn:theta1}
	\theta_{j1}(t_1)\ =\ \frac{\mu^A_{j1}(t_1)}{\mu^S_{j1}(t_1)},
\ee
for which the numerator, termed the \textit{adjusted cumulative readmission rate}, is defined to be:
\[
	\mu^A_{j1}(t_1)\ =\ \frac{1}{n_j}\sum_{i=1}^{n_j} F_{ji1}(t_1; V_{j,1}, V_{j,2}),
\]
where
\[
	 F_{ji1}(t_1; V_{j,1}, V_{j,2})\ =\ \int_0^{t_1} h_1(s| \bfX_{ji1}; \gamma_{ji}, h_{01}, \bfbeta_1, V_{j1}) \exp\left\{-\sum_{g=1}^2 H_g(s| \bfX_{jig}; \gamma_{ji}, h_{0g}, \bfbeta_g, V_{jg})\right\}\ \partial s,
\]
with $H_g(s| \cdot) = \int_0^s h_g(u| \cdot) \partial u$. Note, $F_{ji1}(t_1; V_{j1})$ is the cumulative incidence function for readmission with death taken as a competing risk~\citep{fine1999proportional}. Finally, the denominator in expression (\ref{eqn:theta1}), termed the \textit{standardized adjusted cumulative readmission rate}, is defined to be:
\[
	\mu^S_{j1}(t_1)\ =\ \frac{1}{n_j}\sum_{i=1}^{n_j} \mbox{E}[F_{ji1}(t_1; V_1, V_2)],
\]
where the expectation is with respect to the joint distribution of $(V_1, V_2)$. 

\subsection{Mortality}
\label{sec:measures:T2}

To characterize hospital performance with respect to mortality, we define the \textit{cumulative excess mortality ratio} for $t_2 > 0$ as:
\be
\label{eqn:theta2}
	\theta_{j2}(t_2)\ =\ \frac{\mu^A_{j2}(t_2)}{\mu^S_{j2}(t_2)},
\ee
for which the numerator, termed the \textit{adjusted cumulative mortality rate}, is defined to be:
\[
	\mu^A_{j2}(t_2)\ =\ \frac{1}{n_j}\sum_{i=1}^{n_j} F_{ji2}(t; \bfV_j),
\]
where $F_{ji2}(t; \bfV_j)$ is the CDF for the marginal distribution of $T_2$ for the $i^{th}$ individual in the $j^{th}$ hospital induced by the hierarchical illness-death model, given by:
\be
\label{eqn:F2:V}
	F_{ji2}(t; \bfV_j)\ =\ \int_0^{t_2}\left[\int_0^s f_U(u, s; \bfV_j)\partial u\ +\ f_{\infty}(s; \bfV_j)\right]\partial s,
\ee
where, $f_U(t_1, t_2)$ is the density for the induced joint distribution on the upper wedge of the support of ($T_1, T_2)$ and $f_{\infty}(t_2)$ is the probability mass corresponding to the timing of the terminal event for patients who experience it prior to the non-terminal event~\citep{frydman2010estimation, Xu:Kalb:2010, lee2015bayesian}. Additional details are given in Supplementary Materials Section SM-4. Finally, the denominator of expression (\ref{eqn:theta2}), which we term the \textit{standardized adjusted cumulative mortality rate}, is defined to be:

\[
	\mu^S_{j2}(t_2)\ =\ \frac{1}{n_j}\sum_{i=1}^{n_j} \mbox{E}[F_{ji2}(t; \bfV)],
\]
where the expectation is with respect to the joint distribution of $\bfV$.

\subsection{Interpretation and use}
\label{sec:measures:interpretation}

The interpretations of $\theta_{j1}(t_1)$ and $\theta_{j2}(t_2)$ are analogous to that of $\theta_j$ described in Section \ref{sec:binary}. That is, the two metrics can be interpreted as the extent to which the `observed' readmission and mortality rates for $j^{th}$ hospital differ from the corresponding `expected' rates, for the specific number and case-mix of patients actually treated at the $j^{th}$ hospital. One key difference, however, is that the proposed measures are defined specifically to be functions of the underlying time scales thus providing additional scope and flexibility in the choice of measure on which to base a decision. Depending on the scientific and/or policy goals of the specific analysis, for example, one might choose a specific time, say $(t_1, t_2)$ = (90, 90) at which to evaluate performance, or consider their trajectories over some fixed time interval, say (0,90] days.

\subsection{Characterization of the posterior distribution of $(\theta_{j1}(t_1), \theta_{j2}(t_2))$}
\label{sec:measures:posterior}

The joint posterior distribution of $(\theta_{j1}(t_1), \theta_{j2}(t_2))$ can be readily-characterized through post-processing of the samples generated from the MCMC scheme for the underlying hierarchical semi-competing risks model from Section \ref{sec:clusteredSCR}. One practical challenge is that the integrals in $F_{ji1}(t; V_{j,1}, V_{j,2})$ and $F_{ji2}(t; \bfV_j)$ as well as that expectations in $\mu^S_{j1}(t_1)$ and $\mu^S_{j2}(t_2)$ do not have closed-form expressions. They must, therefore, be evaluated numerically~\citep{abramowitz1964handbook, stoer2013introduction}.

For $F_{ji1}(t; \bfV_j)$ and $F_{ji2}(t; \bfV_j)$, we note that the component integrals are defined over finite intervals (i.e. (0, $t_1$) or (0, $t_2$)). If the analysis of the hierarchical illness-death model has been conducted using parametric specifications for the baseline hazards (e.g. the Weibull distribution; see Section 4.1), so that $F_{ji1}(t; \bfV_j)$ and $F_{ji2}(t; \bfV_j)$ are smooth functions of time then one can use Gauss-Legendre quadrature. Considering $F_{ji2}(t; \bfV_j)$, let $\{(x_{k_1}, w_{k_1}); k_1 = 1, \ldots, K_1)\}$ and $\{(x_{k_2}, w_{k_2}); k_2 = 1, \ldots, K_2)\}$ be the collections of quadrature points and weights based on the Gauss-Legendre rule with $K_1$ and $K_2$ nodes, respectively. These can be obtained, for example, from the \texttt{gaussquad} package in \texttt{R}. Then, for $\tilde{x}_{k_1} = x_{k_1} + 1$ and $\tilde{x}_{k_2} = x_{k_2} + 1$, we have the approximation:
\[
	\widehat{F}_{ji2}(t; \bfV_j)\ =\ \sum_{k_2=1}^{K_2} \frac{tw_{k_2}}{2}\Bigg[\left\{\sum_{k_1=1}^{K_1} \frac{t\tilde{x}_{k_2}w_{k_1}}{4} f_U\left(\frac{t\tilde{x}_{k_1}\tilde{x}_{k_2}}{4}, \frac{t\tilde{x}_{k_2}}{2}; \bfV_j \right)\right\}\ +\ f_{\infty}\left(\frac{t\tilde{x}_{k_2}}{2}; \bfV_j\right) \Bigg].
\]

If the analysis has been conducted using the PEM specification, however, our experience has been that Gauss-Legendre quadrature performs poorly. This may be due, in part, to the non-smooth nature of the induced $F_{ji1}(t; V_{j,1}, V_{j,2})$ and $F_{ji2}(t; \bfV_j)$ at each scan of the MCMC scheme. Since the integration is over a finite interval, however, and that the partition of the interval that underpins the PEM model is known at each scan in the MCMC scheme, one can calculate the relevant integrals exactly.

For $\mu^S_{j1}(t_1)$ and $\mu^S_{j2}(t_2)$, we use Gauss-Hermite quadrature because the integrals within each iteration of the MCMC scheme are with respect to a two- or three-dimensional MVN distribution, respectively, regardless of whether the cluster-specific $\bfV_j$ are taken to arise from a MVN or a Dirichlet process mixture of MVNs (see Section 4.1). Considering, $\mu^S_{j2}(t_2)$, let $\{(x_{k_3}, w_{k_3}); k_3 = 1, \ldots, K_3)\}$, $\{(x_{k_4}, w_{k_4}); k_4 = 1, \ldots, K_4)\}$ and $\{(x_{k_5}, w_{k_5}); k_5 = 1, \ldots, K_5)\}$ be the collections of quadrature points and weights based on the Gauss-Hermite rule with $K_3$, $K_4$ and $K_5$ nodes, respectively. To account for the correlation among the three hospital-specific random effects we use a Cholesky decomposition (i.e. $\Sigma_V = LL^{\top}$) to transform the integrand from one involving uncorrelated variates to one involved correlated variates to give the approximation:
\[
	\widehat{\mbox{E}}[F_{ji2}(t_1; \bfV)]\ =\ \sum_{k_3=1}^{K_3} \sum_{k_4=1}^{K_4} \sum_{k_5=1}^{K_5}\frac{tw_{k_3}w_{k_4}w_{k_5}}{2\pi^{3/2}} \widehat{F}_{ji2}(t; \bfV^L_j)
\]
where $\bfV^L_j = (V^L_{j1}, V^L_{j2}, V^L_{j3})$ with $V^L_{j1} = \sqrt{2}L_{11}x_{k_3}$, $V^L_{j2} = \sqrt{2}(L_{21}x_{k_3}+L_{22}x_{k_4})$ and $V^L_{j3} = \sqrt{2}(L_{31}x_{k_3}+L_{32}x_{k_4}+L_{33}x_{k_5})$.

\subsection{Practical considerations}

As with all use of numerical integration techniques, analysts will need to contend with a trade-off between accuracy and computational burden. For our use of Gaussian quadrature the trade-off is dictated by the number of nodes, together with the number of patients in the sample, $N$=$\sum_{j=1}^Jn_j$ and the number of samples in the MCMC scheme that will be used to characterize the posterior (denoted here by $M$). To calculate the denominators in expression (\ref{eqn:theta1}) across $J$=264 hospitals in the pancreatic cancer data, for example, will require $\approx$4.4$\times$10$^8$ calculations if $M$=1,000 and the number of quadrature nodes is set to five for both $V_1$ and $V_2$. To mitigate the corresponding time burden, the implementation in the \texttt{SemiCompRisks} package for \texttt{R} uses compiled \texttt{C} code as the primary computing engine.

To the best of our knowledge, there are no universal rules regarding the degree of accuracy associated with a given number of nodes. Practically, one strategy is to use some common number of nodes, say $K$, across all instances where a value must be set, and assess the sensitivity of the results as one increases $K$. When the results become insensitive to increases in $K$, one can halt the calculations. For the analyses of the pancreatic cancer data presented in the next subsection, we used this strategy with $K$$\in$$\{3,5,10,15\}$ and found the greatest relative difference between values based on $K$=5 and values based on $K$=15, across all calculations of $\mu^S_{j1}(\cdot)$, $\theta_{j1}(\cdot)$, $\mu^S_{j2}(\cdot)$, and $\theta_{j2}(\cdot)$, to be less an 0.002\%. As such, we present results based on $K$=5.

\subsection{Application to pancreatic cancer data}

Panels (b), (c) and (d) of Figure \ref{fig:PC:joint} provide results based on the PEM-MVN model identified in Section \ref{sec:clusteredSCR:application} as having the best fit to the data; Figures SM-3, SM-4 and SM-5 in the Supplementary Materials provide corresponding results for the other three models. Comparing Figure \ref{fig:PC:joint}(b) with Figure \ref{fig:PC:joint}(a), we find that overall variation across the $J$=264 hospitals in the performance for 90-day readmission is similar under the two analyses, with the posterior medians varying from 0.76 to 1.37 under the semi-competing risks analysis compared to 0.76 to 1.19 under the logistic-Normal. Furthermore, although the ranges differ, the variation in the posterior medians for the hospital-specific excess 90-day mortality ratios is also similar, with those under the semi-competing risks analysis varying between 0.76 to 2.02 and those under the logistic-Normal analysis varying between 0.40 to 1.63.

Notwithstanding the comparable overall variation, however, Figures \ref{fig:PC:joint}(c) and \ref{fig:PC:joint}(d) indicate that the performance measures for individual hospitals can differ substantially: the percent change in the posterior medians for 90-day readmission (from those based on the logistic-Normal) across the hospitals ranges from -22\% to 42\%, while the percent change in the posterior medians for 90-day mortality ranges from -48\% to 173\%. Moreover, the color-coding in Figure \ref{fig:PC:joint}(c) indicates that these differences can have potentially profound effects, with 67 hospitals (color-coded in green) that `benefit' from explicit consideration of mortality in that their excess 90-day readmission ratio is less than 1.0 (i.e. lower than expected) under the semi-competing risks analysis whereas it was greater than 1.0 (i.e. higher than expected) under the logistic-Normal analysis. Similarly, there are 77 hospitals that `loose' in the sense that they are classified as being poor performers under the semi-competing risks analysis whereas they were good performance under the logistic-Normal analysis. From Figure \ref{fig:PC:joint}(d), 55 hospitals that benefit from the semi-competing risks analysis in terms of their performance classification for 90-day mortality while 59 loose.

\begin{figure}[!htbp]
	\centering
	\includegraphics[width=6in]{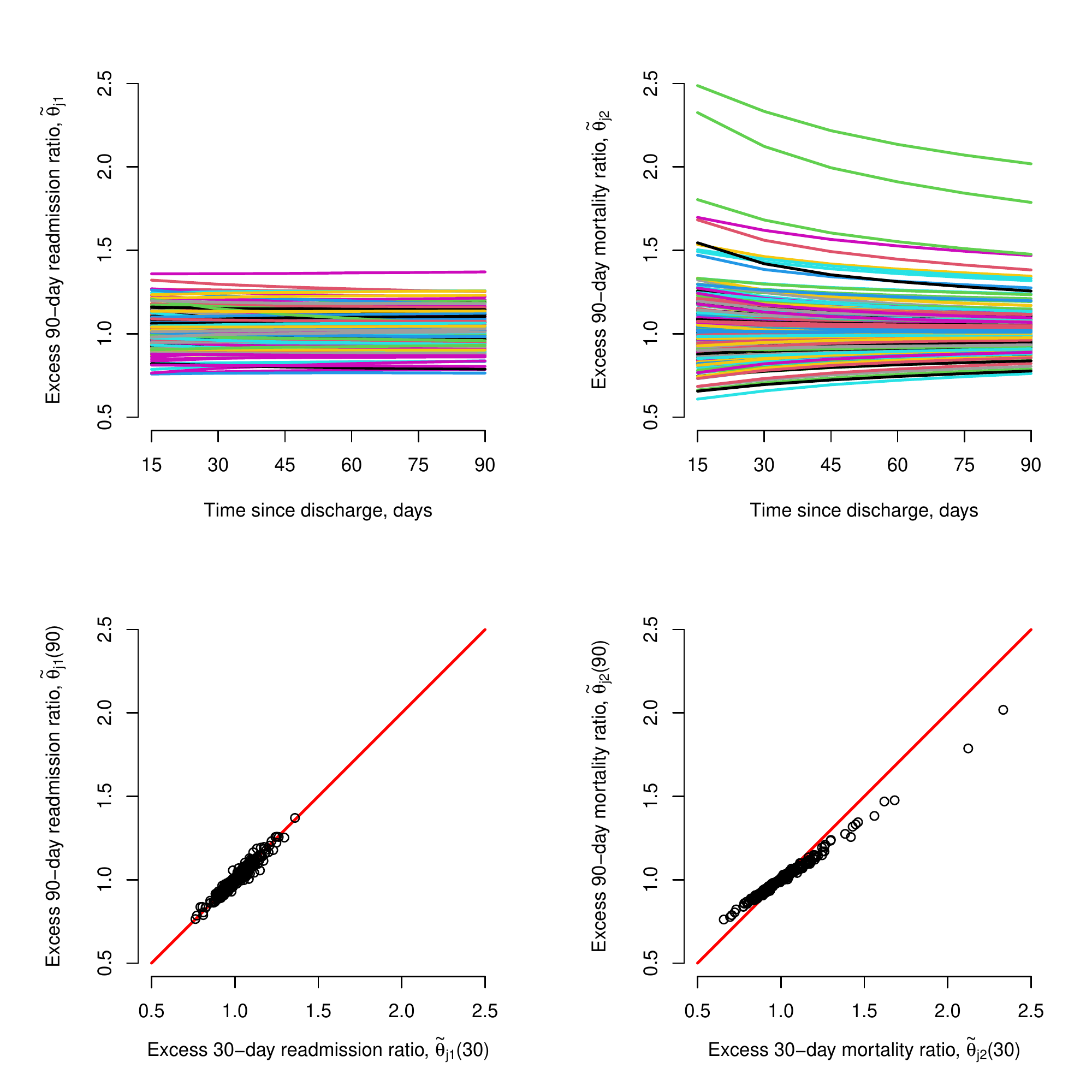}
	\caption{\label{fig:PC:timevarying} Excess readmission and mortality ratios, evaluated at multiple time windows following discharge, across $J$=264 hospitals in California with at least 10 patients aged 65 years or older and diagnosed with pancreatic cancer between 2000-2012.} 
\end{figure}

Finally, while the focus of this case study is on profiling based on outcomes during the 90-day window following discharge, as indicated in Section \ref{sec:measures:interpretation}, the proposed framework permits consideration of other time windows (in particular, without needing to refit the model) and to consider the evolution of $(\theta_{j1}(t_1), \theta_{j2}(t_2))$ over time. To these ends, the top two panels of Figure \ref{fig:PC:timevarying} reports results regarding $(\theta_{j1}(t_1), \theta_{j2}(t_2))$ during six post-discharge windows: 15-, 30-, 45-, 60-, 75- and 90-days. Interestingly, the values for $\theta_{j1}(t_1)$ are fairly stable over time, while those for $\theta_{j2}(t_2)$ seem to attenuate, with less variability across hospitals in the performance metric evaluated over $(0, 90]$-days than over, say, $(0, 30]$-days. This is further clarified in the two lower panels of Figure \ref{fig:PC:timevarying}. See Figures SM-6 through SM-10 in the Supplementary Materials for additional detail comparing the 90-day window to each of the five shorter ones.

\section{Profiling}
\label{sec:goals}

Following the conduct of a Bayesian analyses, it is typical that summaries of the posterior distribution be reported. For example, the posterior mean or the posterior median is often reported as a point estimate of the corresponding parameter. Additionally, the posterior standard deviation or a 95\% credible interval may be reported as a means to communicate uncertainty. In line with this, the results in Figures \ref{fig:PC:observedY} and \ref{fig:PC:joint} regarding excess 90-day readmission and mortality are all based on using the posterior median as a summary for each hospital's performance measure. The use of the posterior median as a point estimate is intuitive but can also be formally justified as being optimal with respect to L$_1$ loss. Whether this is a reasonable thing to do depends, however, on whether the L$_1$ loss is an appropriate loss function for the substantive goal at hand. In this section we consider settings where the substantive goal of the analysis is to profile the performance of the hospitals, possibly towards making some policy decision such as whether to target a hospital with a quality improvement program or whether and how to adjust their reimbursement rates. 

\subsection{Loss function-based profiling}

Let $\bftheta_1 = (\theta_{11}(\cdot), \ldots, \theta_{J1}(\cdot))$ and $\bftheta_2 = (\theta_{12}(\cdot), \ldots, \theta_{J2}(\cdot))$ denote the collections of $J$ hospital-specific cumulative excess readmission and mortality rates, respectively. We define $\Phi_j \equiv \Phi_j(\bftheta_1, \bftheta_2)$ to be some classification function that corresponds to the profiling goal of interest. For example, following \citet{Lin:Loui:2009}, suppose the profiling goal is establish a framework for identifying the top 100($1-\gamma$)\% of hospitals with respect to their performance on the basis of 90-day readmission, so that some action can be taken (e.g. the hospital is rewarded in some way). Addressing this goal could be achieved by ascertaining whether a hospitals' rank is greater than or less than $\gamma(J+1)$, for which $\Phi_j = I\{\mbox{rank}(\theta_{j1}(90)) < \gamma(J+1)\}$, where $I\{\cdot\}$ is an indicator function. As a second example, suppose the goal is to identify whether a hospital has higher-/lower-than-expected for 90-day readmission and mortality (i.e. to classify a hospital into one of the four quadrants in the right-hand panel of Figure \ref{fig:PC:observedY}), and then perform some appropriate action. Towards this, a hospitals' classification could be characterized via:
\be
  \Phi_j\ =\ \begin{cases}
    1 & \text{if $\theta_{j1}(90) > 1.0$ and $\theta_{j2}(90) > 1.0$}, \\
    2 & \text{if $\theta_{j1}(90) > 1.0$ and $\theta_{j2}(90) \leq 1.0$}. \\        
    3 & \text{if $\theta_{j1}(90) \leq 1.0$ and $\theta_{j2}(90) > 1.0$}, \\
    4 & \text{if $\theta_{j1}(90) \leq 1.0$ and $\theta_{j2}(90) \leq 1.0$}, \\
  \end{cases} \nonumber
\ee

Since the true $\bftheta_1$ and $\bftheta_2$ are unknown, however, the true $\Phi_j$ classifications are also unknown and a hospitals' performance must be ascertained on the basis of what we learn from the data. One option for doing this would be to estimate $\Phi_j$ by plugging in the posterior medians of the components of $\bftheta_1$ and $\bftheta_2$ to give $\widetilde{\bfPhi} = (\widetilde{\Phi}_1, \ldots, \widetilde{\Phi}_J)$. While intuitive, the use of the posterior median is not directly motivated by the profiling goal at hand. Indeed it is agnostic to the profiling goal and, as such, could be viewed as an arbitrary choice with no better justification than using, say, the posterior mean. To avoid this arbitrariness, one could ascertain the value of $\Phi_j$ for a given hospital through consideration of a loss function that is tailored specifically to the profiling goal. Let $L(\bfPhi^*; \bfPhi^0)$ denote such a loss function, with $\bfPhi^*$ denoting some candidate value of $\bfPhi$ and $\bfPhi^0$ the true value. Intuitively, $L(\bfPhi^*; \bfPhi^0)$ represents the magnitude of the penalty that one is willing to incur as a result of classifying the $J$ hospitals as $\bfPhi^*$ when the truth is $\bfPhi^0$. 

Returning to the goal of identifying the top 100($1-\gamma$)\% hospitals, \cite{Lin:Loui:2009} consider a number of loss functions that vary in the penalty that is incurred for missclassifying a hospital as being in the 100($1-\gamma$)\% when they are not and/or missclassifying a hospital as not being in the 100($1-\gamma$)\% when they are. One specific option is to assign equal weight (of 1.0) to each such instance of a missclassification (i.e. regardless of the type) to give:
\be
	L(\bfPhi^*; \bfPhi^0)\ =\ \frac{1}{J} \sum_{j=1}^J\ \left[\sum_{c \neq c^\prime} I\{\Phi^0_j = c,\ \Phi^*_j = c^\prime\}\right],
\label{eqn:loss:1}
\ee
where $c$ and $c^\prime$ take on values in $\{0,1\}$, which corresponds to the average number of missclassifications among the $J$ hospitals.

For the second example in which the goal is to categorize hospitals on the basis of whether they have higher-/lower-than-expected rates for 90-day readmission and mortality, one could consider the loss function:
\be
	L(\bfPhi^*; \bfPhi^0)\ =\ \frac{1}{J} \sum_{j=1}^J\ \left(\sum_{c=1}^4 \left[\sum_{c^\prime=1}^4 w(c, c^\prime) I\{\Phi^0_j=c,\ \Phi^*_j=c^\prime\} \right]\right)
\label{eqn:loss:2}
\ee
where $w(c, c^\prime)$ = 0 for $c = c^\prime$ and $w(c, c^\prime)$ for $c \neq c^\prime$ is a penalty that is incurred for classifying a hospital in category $c^\prime$ when the truth is that they are in category $c$. Note, when $w(c, c^\prime)$ = 1.0 for all combinations of $c^\prime$ and $c$, then expression (\ref{eqn:loss:2}) corresponds to the average number of missclassifications among the $J$ hospitals.

\subsection{Estimation}

For a given loss function, an estimate of $\bfPhi$ is obtained by minimizing posterior expected loss or Bayes risk, $\mbox{BR}(\bfPhi^*) = \mbox{E}_\pi[L(\bfPhi^*; \bfPhi)]$ with respect to $\bfPhi^*$. Unfortunately, in all but the most trivial settings $\mbox{BR}(\bfPhi^*)$ will not be analytically tractable because it requires integrating over the full joint posterior distribution of the underlying model specification. As such, it will not be possible to write down a closed-form expression for the corresponding minimizer. To resolve this dilemma, we adopt a strategy in which an estimate is obtained by minimizing with respect to $\bfPhi^*$ an approximation of the Bayes risk, specifically:
\[
	\widehat{\mbox{BR}}(\bfPhi^*)\ =\ \frac{1}{M}\sum_{m=1}^M L(\bfPhi^*; \bfPhi^{(m)}),
\]
where $M$ is the number of samples retained from the MCMC scheme (i.e. after removing burn-out and thinning) and $\bfPhi^{(m)}$ is the value of $\bfPhi$ in the $m^{th}$ such sample.

Operationally, this could be achieved post-model fit (i.e. after all of the MCMC samples have been generated), by: (i) enumerating all possible $\bfPhi^*$, which we denote as $\mathcal{Q}$; (ii) evaluating $\widehat{\mbox{BR}}(\bfPhi^*)$ for all $\bfPhi^* \in \mathcal{Q}$; and, (iii) selecting the $\bfPhi^*$ that corresponds to the smallest $\widehat{\mbox{BR}}(\bfPhi^*)$. For many profiling settings, however, $\mathcal{Q}$ will be massive so that this brute-force strategy will be computationally prohibitive. To see this consider the two profiling goals presented so far. For the first of these, if we say that interest lies in identifying the top 10\% hospitals for the CA data, then $\mathcal{Q}$ consists of ${264 \choose 26} \approx 4.1\times10^{35}$ potential classifications. For the second profiling goal of characterizing a hospitals joint performance status with respect to 90-day readmission and 90-day mortality, then $\mathcal{Q}$ consists of $4^{264} \approx 8.8\times10^{158}$ potential classifications.

To resolve this we propose that researchers adopt one or both of two ad-hoc strategies aimed at reducing overall computational burden. The first strategy is a pre-processing step aimed identifying a subset $\mathcal{Q}_s \subset \mathcal{Q}$ through consideration of $\widetilde{\bfPhi}$, the value of $\bfPhi$ obtained by plugging in the posterior medians of the components of $\bftheta_1$ and $\bftheta_2$. The motivation for doing so is that, although not optimal with respect to the chosen $L(\bfPhi^*; \bfPhi^0)$, $\widetilde{\bfPhi}$ may be a reasonable basis for ruling out select classifications that are \textit{a posteriori} unlikely to be optimal. For example, if a hospital ranks as worst on the basis of the $J$ posterior medians of $\Phi_j = \mbox{rank}(\theta_{j1}(90))$ then it may be reasonable to exclude from $\mathcal{Q}$ any classification that places this hospital in the top 10\%. Similarly, if a hospital has large values of $\widetilde{\theta}_{j1}(90)$ and $\widetilde{\theta}_{j2}(90)$ that are both $>$ 1.0, then it may be reasonable to argue that there is little mass in the joint posterior of ($\theta_{j1}(90), \theta_{j2}(90)$) that supports classifying the hospital in the category where there are lower than expected 90-day rates for both readmission and mortality. Depending on the nature of the approach to ruling out certain classification, the subset $\mathcal{Q}_s$ may be substantially smaller than $\mathcal{Q}$, thus rendering a subsequent brute-force search for the classification that yields the minimum $\widehat{\mbox{BR}}(\bfPhi^*)$.

The second strategy is a sequential updating algorithm aimed at speeding up the task of finding the minimizer from among the classifications in $\mathcal{Q}$ (or $\mathcal{Q}_s$):
\begin{itemize}
	\item[(1)] Let $\widehat{\bfPhi}^{(l)} = (\hat{\phi}_1^{(l)}, \hat{\phi}_2^{(l)}, \ldots, \hat{\phi}_J^{(l)})$ be the current classification of the $J$ hospitals, and $\widehat{\mbox{BR}}(\widehat{\bfPhi}^{(l)})$ the corresponding approximate Bayes risk. 
	\item[(2)] Let $\mathcal{J}$ be some random re-ordering of $\{1, \ldots, J\}$.
	\item[(3)] Let $j^*$ be the first element in $\mathcal{J}$, and consider all possible options for an update of the classification (i.e. $\hat{\phi}_{j^*}^{(l+1)}$). Note, for some loss functions an update for one hospital may require a parallel update to some other hospital. For the loss function given by expression (\ref{eqn:loss:1}), for example, updating a hospital's classification as being in the top 10\% will require moving one hospital that is currently in the top 10\% into the bottom 10\%. We propose that such a hospital be chosen at random, possibly with the acknowledgement of any decisions made through the pre-processing strategy that reduced $\mathcal{Q}$ to $\mathcal{Q}_s$. 
	\item[(4)] For each classification identified in step (3), compute the approximate Bayes risk based on $(\hat{\phi}_1^{(l)}, \ldots, \hat{\phi}_{j^*}^{(l+1)}, \ldots, \hat{\phi}_J^{(l)})$. If the minimum of these values is less than the current Bayes risk then `update' $\hat{\phi}_{j^*}^{(l+1)}$ with the corresponding classification.
	\item[(5)] Repeat steps (3) and (4) for all elements in $\mathcal{J}$.
	\item[(6)] Repeat steps (1)-(5) until no further updates yield smaller values of the approximate Bayes risk (i.e. $\widehat{\mbox{BR}}(\widehat{\bfPhi}^{(l+1)})$ = $\widehat{\mbox{BR}}(\widehat{\bfPhi}^{(l)})$).
\end{itemize}

We note that, although use of either of the proposed strategies is not accompanied by any guarantee that the optimal classification will be obtained, in our experience use of the random re-ordering in step (2) together with a range of starting values (including $\widetilde{\bfPhi}$) invariably led to the same final classification indicating a degree of robustness.

\subsection{Application to pancreatic cancer data}
\label{sec:goals:application}

\subsubsection{Identification of the top 10\% of the hospitals}

Table \ref{tab:classification:L1} summarizes the joint classifications of the $J$=264 hospitals on the basis of whether they ranked in the top 10\% for performance with respect to 90-day readmission and the top 10\% for performance with respect to 90-day mortality. Specifically, for each of the four models fit in Section \ref{sec:clusteredSCR:application}, the sub-tables in Table \ref{tab:classification:L1} present a classification based on using the posterior median for $(\theta_{j1}(90), \theta_{j2}(90))$ as well as that based on minimizes the Bayes risk with loss function (\ref{eqn:loss:1}). Inspection of the margins of the four tables indicates that the number of hospitals that end up in one of the four categories is roughly the same whether the classification is based on the posterior medians or via minimization of the Bayes risk. Furthermore, the marginal distributions are relatively robust across the four models. However, the cross-classification between the results based on the posterior median and those based on minimizing the Bayes risk reveals that there is some discordance between the two approaches. Under the PEM-MVN model, for example, 7 hospitals that are classified as not being in the top 10\% for either outcome when one uses the posterior median are classified as being a top performer for one of them when one minimizes the Bayes risk. Again, however, these observations seem to hold across the four fitted models.

\begin{table}
\caption{\label{tab:classification:L1}Classification of $J$=264 hospitals according to whether they are ranked in the top 10\% for 90-day readmission and in the top 10\% for 90-day mortality on the basis of: (i) the posterior median for $(\theta_{j1}(90), \theta_{j2}(90))$; or, (ii) the minimizer of the approximate Bayes risk, $\widehat{\mbox{BR}}(\bfPhi^*)$, based on the loss function given by expression (\ref{eqn:loss:1}).}
\centering
\scalebox{1.1}{
\begin{tabular}{c c ccccc}
\\
\multicolumn{1}{l}{\fbox{\bf{WB-MVN}}}				
						& 			& \multicolumn{5}{c}{\underline{\textbf{Loss function-based}}} \\
						& 			& No/No & No/Yes & Yes/No & Yes/Yes & \\ \cline{3-6}
						& No/No		& \mc{203} & \mc{4} & \mc{3} & \mc{0} & 210 \\ \cline{3-6}
\underline{\textbf{Posterior}}	& No/Yes		& \mc{4} & \mc{23} & \mc{0} & \mc{0} & 27 \\ \cline{3-6}
\underline{\textbf{median}}	& Yes/No		& \mc{3} & \mc{0} & \mc{24} & \mc{0} & 27 \\ \cline{3-6}
						& Yes/Yes		& \mc{0} & \mc{0} & \mc{0} & \mc{0} & 0 \\  \cline{3-6}
						& 			& 210 & 27 & 27 & 0 &  \\
\\
\multicolumn{1}{l}{\fbox{\bf{WB-DPM}}}				
						& 			& \multicolumn{5}{c}{\underline{\textbf{Loss function-based}}} \\
						& 			& No/No & No/Yes & Yes/No & Yes/Yes & \\ \cline{3-6}
						& No/No		& \mc{203} & \mc{3} & \mc{4} & \mc{0} & 210 \\ \cline{3-6}
\underline{\textbf{Posterior}}	& No/Yes		& \mc{3} & \mc{24} & \mc{0} & \mc{0} & 27 \\ \cline{3-6}
\underline{\textbf{median}}	& Yes/No		& \mc{4} & \mc{0} & \mc{23} & \mc{0} & 27 \\ \cline{3-6}
						& Yes/Yes		& \mc{0} & \mc{0} & \mc{0} & \mc{0} & 0 \\  \cline{3-6}
						& 			& 210 & 27 & 27 & 0 &  \\
\\
\multicolumn{1}{l}{\fbox{\bf{PEM-MVN}}}				
						& 			& \multicolumn{5}{c}{\underline{\textbf{Loss function-based}}} \\
						& 			& No/No & No/Yes & Yes/No & Yes/Yes & \\ \cline{3-6}
						& No/No		& \mc{204} & \mc{3} & \mc{4} & \mc{0} & 211 \\ \cline{3-6}
\underline{\textbf{Posterior}}	& No/Yes		& \mc{3} & \mc{23} & \mc{0} & \mc{0} & 26 \\ \cline{3-6}
\underline{\textbf{median}}	& Yes/No		& \mc{4} & \mc{0} & \mc{22} & \mc{0} & 26 \\ \cline{3-6}
						& Yes/Yes		& \mc{0} & \mc{0} & \mc{0} & \mc{1} & 1 \\  \cline{3-6}
						& 			& 211 & 26 & 26 & 1 &  \\
\\
\multicolumn{1}{l}{\fbox{\bf{PEM-DPM}}}				
						& 			& \multicolumn{5}{c}{\underline{\textbf{Loss function-based}}} \\
						& 			& No/No & No/Yes & Yes/No & Yes/Yes & \\ \cline{3-6}
						& No/No		& \mc{204} & \mc{3} & \mc{4} & \mc{0} & 211 \\ \cline{3-6}
\underline{\textbf{Posterior}}	& No/Yes		& \mc{3} & \mc{22} & \mc{0} & \mc{0} & 25 \\ \cline{3-6}
\underline{\textbf{median}}	& Yes/No		& \mc{5} & \mc{0} & \mc{21} & \mc{0} & 26 \\ \cline{3-6}
						& Yes/Yes		& \mc{0} & \mc{0} & \mc{0} & \mc{1} & 1 \\  \cline{3-6}
						& 			& 212 & 25 & 25 & 1 &  \\
\end{tabular}}
\end{table}

\subsubsection{Bivariate classification}

Table \ref{tab:classification:L2} summarizes the joint classifications of the $J$=264 hospitals on the basis of whether they are found to have higher- or lower-than expected 90-day readmission and 90-day mortality. Specifically, for each of the four models fit in Section \ref{sec:clusteredSCR:application}, the sub-tables in Table \ref{tab:classification:L2} present a classification based on using the posterior median for $(\theta_{j1}(90), \theta_{j2}(90))$ as well as that based on minimizing the Bayes risk with loss function (\ref{eqn:loss:1}).

Focusing on the results based on the PEM-MVN model, we find that 6 hospitals initially are classified as being good performers for both 90-day readmission and 90-day mortality (i.e. in the `Lower/Lower' category) when the posterior medians are used are reclassified as being poor performers for at least one of the outcomes: 4 are reclassified as being poor performers with respect to 90-day readmission and 2 with respect to 90-day mortality. Furthermore, 5 hospitals that are initially classified as being poor performers for both outcomes (i.e. in the `Upper/Upper' category) are classified as being good performers for at least one outcome: 3 are reclassified as being good performers with respect to 90-day mortality and 2 with respect to 90-day readmission.

Looking at the loss-function based classifications across the four models we see that there are some important differences. For example, while 39 hospitals are classified as being poor performers for both 90-day readmission and 90-day mortality under the PEM-MVN model, this number increases to 58 under the PEM-DPM model. Similarly, while 37 hospitals are classified as being good performers for both 90-day readmission and 90-day mortality under the PEM-MVN model, this number decreases to 27 under the PEM-DPM model.

\begin{table}
\caption{\label{tab:classification:L2}Classification of $J$=264 hospitals according to whether they are found to have higher- or lower-than expected 90-day readmission and 90-day mortality on the basis of: (i) the posterior median for $(\theta_{j1}(90), \theta_{j2}(90))$; or, (ii) the minimizer of the approximate Bayes risk, $\widehat{\mbox{BR}}(\bfPhi^*)$, based on the loss function given by expression (\ref{eqn:loss:2}).}
\centering
\scalebox{1.1}{
\begin{tabular}{c c ccccc}
\\
\multicolumn{1}{l}{\fbox{\bf{WB-MVN}}}				
						&				& \multicolumn{5}{c}{\underline{\textbf{Loss function-based}}} \\
						&				& Higher/ & Higher/ & Lower/ & Lower/ & \\
						&				& Higher & Lower & Higher & Lower & \\ \cline{3-6}
						& Higher/Higher	& \mc{27} & \mc{3} & \mc{4} & \mc{0} & 34 \\ \cline{3-6}
\underline{\textbf{Posterior}}	& Higher/Lower		& \mc{0} & \mc{106} & \mc{0} & \mc{0} & 106 \\ \cline{3-6}
\underline{\textbf{median}}	& Lower/Higher		& \mc{0} & \mc{0} & \mc{90} & \mc{0} & 90 \\ \cline{3-6}
						& Lower/Lower 		& \mc{0} & \mc{3} & \mc{4} & \mc{27} & 34 \\  \cline{3-6}
						&				& 27 & 112 & 98 & 27 &  \\
\\
\multicolumn{1}{l}{\fbox{\bf{WB-DPM}}}
						&				& \multicolumn{5}{c}{\underline{\textbf{Loss function-based}}} \\
						&				& Higher/ & Higher/ & Lower/ & Lower/ & \\
						&				& Higher & Lower & Higher & Lower & \\ \cline{3-6}
						& Higher/Higher	& \mc{20} & \mc{4} & \mc{3} & \mc{0} & 27 \\ \cline{3-6}
\underline{\textbf{Posterior}}	& Higher/Lower		& \mc{0} & \mc{112} & \mc{0} & \mc{0} & 112 \\ \cline{3-6}
\underline{\textbf{median}}	& Lower/Higher		& \mc{0} & \mc{0} & \mc{84} & \mc{0} & 84 \\ \cline{3-6}
						& Lower/Lower 		& \mc{0} & \mc{5} & \mc{5} & \mc{31} & 41 \\  \cline{3-6}
						&				& 20 & 121 & 92 & 31 &  \\
\\
\multicolumn{1}{l}{\fbox{\bf{PEM-MVN}}}				
						&				& \multicolumn{5}{c}{\underline{\textbf{Loss function-based}}} \\
						&				& Higher/ & Higher/ & Lower/ & Lower/ & \\
						&				& Higher & Lower & Higher & Lower & \\ \cline{3-6}
						& Higher/Higher	& \mc{39} & \mc{3} & \mc{2} & \mc{0} & 44 \\ \cline{3-6}
\underline{\textbf{Posterior}}	& Higher/Lower		& \mc{0} & \mc{97} & \mc{0} & \mc{0} & 97 \\ \cline{3-6}
\underline{\textbf{median}}	& Lower/Higher		& \mc{0} & \mc{0} & \mc{80} & \mc{0} & 80 \\ \cline{3-6}
						& Lower/Lower 		& \mc{0} & \mc{4} & \mc{2} & \mc{37} & 43 \\  \cline{3-6}
						&				& 39 & 104 & 84 & 37 &  \\
\\
\multicolumn{1}{l}{\fbox{\bf{PEM-DPM}}}				
						&				& \multicolumn{5}{c}{\underline{\textbf{Loss function-based}}} \\
						&				& Higher/ & Higher/ & Lower/ & Lower/ & \\
						&				& Higher & Lower & Higher & Lower & \\ \cline{3-6}
						& Higher/Higher	& \mc{58} & \mc{4} & \mc{3} & \mc{0} & 65 \\ \cline{3-6}
\underline{\textbf{Posterior}}	& Higher/Lower		& \mc{0} & \mc{79} & \mc{0} & \mc{0} & 79 \\ \cline{3-6}
\underline{\textbf{median}}	& Lower/Higher		& \mc{1} & \mc{0} & \mc{84} & \mc{0} & 85 \\ \cline{3-6}
						& Lower/Lower 		& \mc{0} & \mc{6} & \mc{2} & \mc{27} & 35 \\  \cline{3-6}
						&				& 59 & 89 & 89 & 27 &  \\
\end{tabular}}
\end{table}

\section{Discussion}
\label{sec:discussion}

The current statistical paradigm for quantifying hospital performance with respect to readmission ignores death as a competing risk. Doing so may, arguably, be reasonable for health conditions with low mortality, although it is unclear what threshold for `low' should be used or even whether a single such threshold exists. Either way, ignoring death as a competing risk is unlikely to be reasonable for monitoring performance corresponding to terminal conditions for which the clinical focus is often on managing end-of-life quality of care for the patient. For such conditions, given the substantial financial implications involved, quantifying the magnitude of variation and the attributes of hospitals associated with performance on both readmission and mortality in parallel could help to direct and prioritize quality improvement initiatives. This, we argue, represents a fundamental shift away from uni-dimensional assessments of readmission-based performance for individual health conditions as well as hospital-wide assessments that consider a range of conditions simultaneously~\citep{CMS2019HWR}. Practically, to achieve this, we have proposed a general framework consisting of four components: (i) a hierarchical model for the underlying semi-competing risks data; (ii) two novel measures of performance; (iii) a loss-function based approach to classifying the performance of a collection of hospitals; and, (iv) a series of pragmatic strategies for mitigating computational burden. While the first of these was described and evaluated via simulation in \cite{lee2016hierarchical}, components (ii)-(iv) are the key contributions of this paper.

The proposed framework is motivating by an on-going collaboration regarding end-of-life care for patients with a terminal cancer diagnosis. Within this backdrop, this paper presents key methodologic issues and the proposed framework via a detailed case-study using data on $N$=17,685 patients diagnosed with pancreatic cancer at one of $J$=264 hospitals in California between 2000-2012. Several interesting aspects of the case study deserve additional discussion. First, from a substantive perspective, the results show that the classification of a given hospital may change when one grounds the evaluation of performance within the semi-competing risks framework in lieu of using the output from a logistic-Normal GLMM, and when one uses a loss-function based approach to classifying hospitals in lieu of using the posterior median as a plug-in estimator. In considering the differences between the results based on a logistic-Normal model and those based on a hierarchical model for semi-competing risks (see Figures \ref{fig:PC:joint}(c) and \ref{fig:PC:joint}(d)), additional analyses reveal that hospital volume seemed to play little-to-no systematic role in dictating whether a hospitals was reclassified on the basis of readmission. However, mortality prior to readmission did appear to play a meaningful role. Specifically, in regard to the loss function given by expression (\ref{eqn:loss:1}) we found that hospitals with (relatively) low or high mortality prior to readmission were more likely to be classified as `winners' (i.e. they benefitted from the use of the semi-competing risks analysis). Furthermore, in regard to the loss function given by expression (\ref{eqn:loss:2}) we found that the rate at which hospitals were differentially being reclassified as winners (i.e. being reclassified into the `lower-than-expected' group) increased with the mortality rate prior to readmission. Whether these observations/insight are generalizable to all settings is unclear, however; whether differences manifest is likely a function of many aspects of the data (e.g. the patient and hospital characteristics) and underlying covariation in the outcomes.

A second interesting aspect of the case study is the sensitivity of the classifications across the four model specifications in Table \ref{tab:classification:L2}. In considering the relatively complex nature of cluster-correlated semi-competing risks data, that this is the case should not, we believe, be unexpected. Indeed, given that the PEM specification for the baseline hazards is substantially more flexible than the Weibull specification and, similarly, that the DPM specification for the hospital-specific random effects is substantially more flexible than the MVN, it would be surprising if there was no variability in the results across the specifications. Moreover, even in the more standard binary outcome setting, the use of a MVN for the random effects versus a DPM will likely result in differences in the final fit and hence the final classification(s). Indeed, this is a phenomenon that likely affects all statistical analysis of real data that involve complex, hierarchical structure. Of course, in practice one never knows the `truth', so that the use of established and understood model fit criteria, such the DIC and/or LMPL measures referenced in Section \ref{sec:clusteredSCR:analyses}, are critical tools, specifically as a means to choosing the final model \textit{before} seeing the final profiling results.

Finally, the results in Figure \ref{fig:PC:timevarying}, together with those in Figures SM-6 through SM-10 of the Supplementary Materials, serve to highlight the policy opportunities associated with embedding the task at-hand within a hierarchical semi-competing risks model. Specifically that the model explicitly considers and borrows across time provides a framing for informing policy interventions that are tailored to, for example, improving performance in the immediate aftermath following discharge without affecting longer-term performance. In considering this, it is important to note that the trajectories in Figure \ref{fig:PC:timevarying} may be (generally) parallel, in part at least, because of the sole inclusion of random intercepts in models (\ref{HID:h1})-(\ref{HID:h3}). One way to enhance this model would be to additionally include random slopes for time, although the interpretation of what these components capture would need to be carefully considered in the profiling context of this paper. Finally, in considering whether and how current federal programs may be expanded to include end-of-life conditions such as pancreatic cancer, results such as those presenting in Figure \ref{fig:PC:timevarying} may serve as important information in selecting an appropriate window.

The specific profiling context that motivated this work is that of how federal agencies in the United States incentivize quality improvement in relation to all-cause readmission, specifically through the Hospital Inpatient Quality Improvement Program and the Hospital Readmissions Reduction Program. Outside of this context, the identification of ``unusual'' health-care providers is also an important policy goal, one for which an assessment of whether a performance metric, such as $\theta_{j1}(t_1)$ or $\theta_{j2}(t_2)$, lies above or below some threshold (e.g. 1.0) is arguably inappropriate~\citep{jones2011identification}. Moreover, extending methods geared towards this goal, such as the funnel plots of \cite{spiegelhalter2005funnel}, to jointly consider readmission and mortality is an important avenue for future work.

Finally, in considering the role that mortality plays as a competing risk, we have adopted an overarching approach to hospital performance that mirrors the GLMM-based approach originally proposed by~\cite{normand1997statistical} for univariate binary outcomes. For the latter, a number of recent papers have queried various aspects of this approach, advocating for the use of direct standardization to some common population, instead of standardization to the patients actually seen at the hospital~\citep{varewyck2014shrinkage, george2017mortality}; highlighting concern regarding the use of random effects ~\citep{varewyck2014shrinkage, kalbfleisch2013monitoring}; and, arguing for the inclusion of hospital-level characteristics as part of the random effects specification~\citep{silber2010hospital, george2017mortality}. These are all complex issues for which a detailed discussion in the end-of-life profiling context is beyond the scope of this paper. We do note, however, that in mirroring the approach of ~\cite{normand1997statistical}, the proposed methods are aligned with the philosophy that underpins the Hospital Inpatient Quality Improvement Program and the Hospital Readmissions Reduction Program as they are currently run. In that sense, the methods we propose could, in principle, be implemented immediately and provide a means to enhancing quality of end-of-life care.

\section*{Acknowledgements}
This work was supported by NIH grant R01CA181360.

\begin{supplement}
~\stitle{Section SM-1}
\sdescription{Table summarizing patient characteristics and outcomes}
\end{supplement}

\begin{supplement}
\stitle{Section SM-2}
\sdescription{Exploratory data analyses for the CMS pancreatic cancer data}
\end{supplement}

\begin{supplement}
\stitle{Section SM-3}
\sdescription{Details and results from fits of logistic-Normal generalized linear mixed models for the binary outcomes of 90-day readmission and 90-day mortality}
\end{supplement}

\begin{supplement}
\stitle{Section SM-4}
\sdescription{Details and results from the fit of four hierarchical semi-competing risks models to the CMS pancreatic cancer data}
\end{supplement}

\begin{supplement}
\stitle{Section SM-5}
\sdescription{Additional results regarding the performance metrics for the CMS pancreatic cancer data}
\end{supplement}

\begin{supplement}
\stitle{Section SM-6}
\sdescription{Additional technical details regarding $F_{ji2}(t_2; \bfV_j)$}
\end{supplement}


\bibliographystyle{imsart-nameyear} 
\bibliography{ref.bcscr}       

\end{document}